\documentclass[11pt]{article}
\usepackage[totalwidth=450pt,totalheight=584pt]{geometry}
\usepackage{amsmath,amssymb,epsf,graphics,graphicx}
\usepackage{psfrag}
\usepackage{amsfonts}
\usepackage{mathrsfs}
\usepackage{epsfig}
\usepackage{relsize}
\usepackage{epstopdf}
\def\sl(2){\alg{sl}(2)}

\def\a {\alpha}

\def\la{\label}

\def\tp{{\widetilde p}}

\newcommand{\alg}[1]{\mathfrak{#1}}
\newcommand{\CQ}{{\cal Q}}

\newcommand{\vep}{\varepsilon}

\newcommand{\atopfrac}[2]{\genfrac{}{}{0pt}{}{#1}{#2}}

\newcommand{\bem}{\left (\begin{matrix}}
\newcommand{\eem}{\end{matrix} \right )}

\newcommand{\nn}{\nonumber}

\newcommand{\be}{\begin{equation}}
\newcommand{\ee}{\end{equation}}
\newcommand{\ba}{\begin{eqnarray}}
\newcommand{\ea}{\end{eqnarray}}
\newcommand\RR{\mathbb{R}}

\newcommand{\ZZ}{\mathbb{Z}}

\newcommand\blank[1]{}

\newcommand{\fract}[2]{{\textstyle\frac{#1}{#2}}}

\newcommand\eq{\begin{equation}}
\newcommand\en{\end{equation}}
\newcommand\bea{\begin{eqnarray}}
\newcommand\eea{\end{eqnarray}}
\newcommand{\resection}[1]{\setcounter{equation}{0}\section{#1}}

\newcommand\ep{\epsilon}

\begin{document}
\sloppy
\renewcommand{\thefootnote}{\fnsymbol{footnote}}
\newpage
\begin{titlepage}
\vskip 1.2cm
\begin{center}
{\Large{\bf TBA and Y-system for planar $\text{AdS}_4/\text{CFT}_3$}}
\end{center}
\vspace{1.8cm}

\centerline{{\large Diego Bombardelli$^a$,
Davide Fioravanti$^a$ and Roberto Tateo$^b$\footnote{e-mails: bombardelli@bo.infn.it, fioravanti@bo.infn.it, tateo@to.infn.it.}}}
\vskip 0.9cm

\centerline{${}^a$ \small INFN-Bologna and Dipartimento di Fisica,
Universit\`a di Bologna,
}
\centerline{\small Via Irnerio 46, 40126 Bologna, Italy.}
\vskip 0.2cm
\centerline{${}^{b}$ \small Dip.\ di Fisica Teorica and INFN,
Universit\`a di Torino,}
\centerline{\small Via P.\ Giuria 1, 10125 Torino, Italy.}
\vskip 0.2cm

\vskip 1.25cm

\begin{abstract}
\noindent
We conjecture the set of asymptotic Bethe Ansatz equations for the {\it mirror} model of the $\text{AdS}_4\times\mathbb{CP}^3$ string theory, corresponding to the planar $\mathcal{N}=6$ superconformal Chern-Simons gauge theory in three dimensions. Hence, we derive the (vacuum energy) thermodynamic Bethe Ansatz equations and the Y-system  describing the {\it direct} $\text{AdS}_4/\text{CFT}_3$ string theory.

\end{abstract}
\end{titlepage}

\renewcommand{\thefootnote}{\arabic{footnote}}
\setcounter{footnote}{0}

\section{Introductory remarks}

The $\text{AdS}_4/\text{CFT}_3$ strong/weak coupling duality relates conjecturally the type IIA superstring theory on curved space-time $\text{AdS}_4\times\mathbb{CP}^3$ to its boundary theory, namely the conformal ${\cal N}=6$ super Chern-Simons (SCS) gauge theory in three dimensions \cite{Aharony:2008ug}. Similarly to the previously discovered   $\text{AdS}_5/\text{CFT}_4$ correspondence \cite{M-GKP-W}, the initial hints of integrability both in perturbative calculations  of anomalous dimensions (of single trace composite operators in the $SU(2)\times SU(2)$ sector) of SCS  \cite{Minahan:2008hf} and in energies of string configurations on the $\text{AdS}_{4}\times\mathbb{CP}^{3}$ background \cite{Arutyunov:2008if} have clearly marked the beginning of a new and fascinating research topic. In fact, integrability structures and tools typically imply  important enhancements of non-perturbative or exact calculations.

By mimicking the  $\text{SYM}_4$ results and Dynkin diagram structure \cite{BS}, a set of all-loop Bethe Ansatz (BA) equations  valid in the asymptotic regime of large quantum numbers (e.g. size or R-charge $L$, Lorentz spin, etc.) has been proposed for all the SCS sectors \cite{Gromov:2008qe}.  In \cite{Ahn:2008aa} a scattering matrix was devised in such a way that the corresponding  Bethe-Yang equations match the BA equations of \cite{Gromov:2008qe}. For the particular interest of the present work, the finite size corrections implied by this scattering matrix have been investigated and compared \cite{Bombardelli:2008qd} with string theory results \cite{Grignani:2008te}, going  beyond the asymptotic regime and  confirming the {\it bona fide} feature of the S-matrix. Yet, for {\it small} size $L$ the L\"uscher method does not  allow a feasible extension to all loops and thus turns out to be unfit for a large class of interesting operators.

On the contrary, the Thermodynamic Bethe Ansatz (TBA) idea recently succeeded in providing a set of infinite integral equations, which should govern {\it in principle} the spectrum of anomalous dimensions non-perturbatively and for general values of the quantum numbers~\cite{Bombardelli:2009ns,Gromov:2009bc, Arutyunov:2009ur}. The statement `{\it in principle}' results from the consideration that the TBA procedure involves a minimisation to the ground state energy, although  the first few excited states may be extracted from it. In fact, the TBA method for calculating the free energy of statistical field theories at temperature $T$ dates back to Yang and Yang~\cite{YY}, and provides for a minimisation procedure  for both Fermi and Bose statistics \cite{YY}. Furthermore, the idea of combining the TBA with the modular transformation (or double Wick rotation) exchanging space and time in quantum integrable 2D relativistic massive field theories belongs to Al.B. Zamolodchikov~\cite{Zamolodchikov:1989cf}.

The TBA procedure may be briefly described in the following way. Let us assume that the
original ({\it direct}) theory is defined on a torus space-time geometry, the space is a ring of finite circumference $L$, while time runs over a very long circumference $R\rightarrow\infty$. Modular transformation amounts in exchanging space and time, thus defining a mirror theory living on a  space segment of length  $R\rightarrow\infty$. For this theory  the asymptotic  {\it mirror} Bethe-Yang equations are exact. Time is compactified on a circumference $L$ that can be  interpreted as the inverse of the temperature  $T=1/L$ and the Yang-Yang thermodynamic Bethe Ansatz procedure \cite{YY} can be used to find the minimum free energy of the mirror theory or equivalently  the vacuum  energy of the original {\it direct} theory confined on a ring of circumference $L$. In relativistic quantum field theories the ground state TBA equations have been generalised by \cite{Bazhanov:1996aq, Dorey:1996re} to  excited states.

Starting from the equations for the ground state, excited states were obtained in \cite{Dorey:1996re} through a process of analytic continuation in the particle masses and by considering the points where  the singularities of the integrand cross the  integration contour.

The possibility to extend the procedure of~ \cite{Bazhanov:1996aq, Dorey:1996re} to the non-relativistic $\text{AdS}_5\times\text{S}^5$ model was anticipated in the `Partial conclusions and remarks' section in \cite{Bombardelli:2009ns} by emphasising the appearance of new driving terms of the form $\sum_i \ln S(u_i,u)$ as residues of the convolution integrals, with $S$ suitable scattering matrix elements\footnote{This mechanism becomes even simpler and straightforward in the (different) NLIE set-up when starting from the microscopic Bethe Ansatz description \cite{Fioravanti:1996rz} (cf. also \cite{Fioravanti:2007un} about the SYM$_4$ non-relativistic (but asymptotic) case).}.
 This extension was studied in more details   for the excited states of the $sl(2)$ sector in SYM$_4$ in two different articles, \cite{Gromov:2009bc} and \cite{Arutyunov:2009ax}.    In the latter reference some problems and modifications of the equations when varying the 't Hooft coupling $g$ have been highlighted, thus making the problem still more puzzling.

In the context of the  $\text{AdS}_{5}/\text{CFT}_{4}$ correspondence, the mirror theory was brilliantly introduced and analysed in \cite{Arutyunov:2007tc} on the basis of previous  studies on the analytic properties of the direct S-matrix (like for instance \cite{Janik:2006dc, DHM, Roiban:2006gs, Chen:2006gq}). Recently, the  string hypothesis, which is the basis for the TBA procedure, was proposed and studied within the large-size thermodynamic limit \cite{Arutyunov:2009zu}.

Here, bearing this in mind,  we conjecture a set of Bethe Ansatz equations for the mirror theory of $\text{AdS}_{4}/\text{CFT}_{3}$. The latter show bound states of particles of type A with particles of type B
in their $sl(2|1)$ grading. Therefore, we find profitable to formulate the string hypothesis in this grading, though it differs neatly from that proposed in  \cite{Arutyunov:2009zu}. Eventually, we implement the string hypothesis to derive the ground state thermodynamic Bethe Ansatz equations (for the TBA variables or {\it pseudoenergies}). From these equations we also deduce the Y-system, namely {\it universal} functional equations among the  Y-functions, which are nothing but the exponential of the pseudoenergies. Although it is evident that we shall suffer an information loss through this step, the Y-system is believed to be universal in the sense of remaining  the same for excited state equations as well.

The rest of this paper is organised as follows. Section \ref{section2} contains a brief summary on the
magnon dispersion relations for the string $\text{AdS}_{4}/\text{CFT}_{3}$ and its mirror theory (also with the uniformisation variable). The proposed  all-loop  Bethe Ansatz equations for the mirror theory and the corresponding string hypothesis in the $sl(2|1)$  grading are discussed in sections \ref{section3}, \ref{section4} and \ref{section5}. The thermodynamic Bethe Ansatz procedure is implemented
 and the corresponding Y-system derived respectively in  sections \ref{section6} and \ref{section7}.
Section \ref{section8} contains some general  concluding remarks and a short  discussion on the  connection between our Y-system and that proposed, for the same theory, in  \cite{GKVI}.
Finally, in Appendix~A we summarise the S-matrix elements appearing in the Bethe Ansatz equations and defining the TBA kernels, together with the   functional identities enjoyed by the latter.

\resection{Dispersion relations}
\label{section2}
The dispersion relation of the $\text{AdS}_4\times\mathbb{CP}^3$ string theory concerns two species, $A$ and $B$, of  `magnon' excitations  with energy and momentum $(H^\alpha, p^\alpha)$  $(\alpha=A,B)$ related by
\be
H^\a=\frac{1}{2}\sqrt{1+16h^2(\lambda)\sin^2\frac{p^\a}{2}}~,
\label{disprel}
\ee
and total energy $H=\sum_{\a=A,B}H^\a$. Upon the analytic continuation $p^\a\rightarrow i \tilde{H}^\a$ and $H^\a\rightarrow i \tilde{p}^\a$ we obtain the mirror theory dispersion relations
\be
\tilde{H}^\a=2\,\mbox{arcsinh}\frac{\sqrt{1+4(\tilde{p}^\a)^2}}{4h(\lambda)}~.
\label{mirrordisprel}
\ee
The dispersion relation (\ref{disprel}) can be uniformised in terms of Jacobi elliptic functions:

\be
p=2 \mbox{am} (z,k)\,, \quad \sin \frac{p}{2}=\mbox{sn}(z,k)\,,\quad H=\frac{\mbox{dn}(z,k)}{2}\,,
\ee
with $k=-16h^{2}(\lambda)$,  and then parameterised by an elliptic curve with periods

\be\nonumber
2\omega_1=4{\rm K}(k)\, , ~~~~~~~~~ 2\omega_2=4i{\rm K}(1-k)-4{\rm
K}(k)~.
 \ee
For a real particle with momentum $p$ we know that
\be
p=2i\,\mbox{arcsinh}\frac{\sqrt{1+4(\tilde{p}^\a)^2}}{4h(\lambda)}=2 \mbox{am} (z,k)
\ee
and therefore
\be
\tilde{p}=-\frac{i}{2}\mbox{dn}(z,k)~,
\label{pmir}
\ee
that is real only if we shift the variable z by a quarter of the imaginary period.
Thus, if we shift $H$ by $\omega_{2}/2$, then $H\rightarrow i\tilde{p}$, while $p$ becomes

\be
p=2\mbox{am}z\rightarrow 2i\mbox{arccoth}\frac{\sqrt{1-k}}{\mbox{dn}z}=i\tilde{H}~.
\ee
In fact,

\be
2\,\mbox{arccoth}\frac{\sqrt{1-k}}{\mbox{dn}z}=2\,\mbox{arcsinh}\frac{\sqrt{1+4(\tilde{p}^\a)^2}}{4h(\lambda)}
\ee
gives the same expression of (\ref{mirrordisprel}).

\resection{The scattering matrix}
\label{section3}
The $\text{AdS}_{4}/\text{CFT}_{3}$ $S$-matrix was proposed in \cite{Ahn:2008aa}. As there are two sets of momentum carrying excitations, called $A$- and $B$-particles, each of which form a four-dimensional representation of $SU(2|2)$, the conjectured exact $S$-matrix has the following factorised structure:

\ba
S_{AA}(p_{1},p_{2})=S_{BB}(p_{1},p_{2})=S_{0}(p_{1},p_{2})S(p_{1},p_{2})~,\nonumber\\
S_{AB}(p_{1},p_{2})=S_{BA}(p_{1},p_{2})=\tilde{S}_{0}(p_{1},p_{2})S(p_{1},p_{2})~,
\label{S-matrices}
\ea
where $S$ is the $SU(2|2)$ $\text{AdS}_{5}\times S^{5}$ string $S$-matrix \cite{Be, Arutyunov:2006yd} and

\ba
S_{0}(p_{1}\,, p_{2}) &=&
\frac{1-\frac{1}{x^{+}_{1}x^{-}_{2}}}{1-\frac{1}{x^{-}_{1}x^{+}_{2}}}
\sigma(p_{1}\,, p_{2}) \,, \nonumber \\
\tilde S_{0}(p_{1}\,, p_{2}) &=&
\frac{x^{-}_{1}-x^{+}_{2}}{x^{+}_{1}-x^{-}_{2}}
\sigma(p_{1}\,, p_{2}) \,.
\label{S0values}
\ea
In (\ref{S0values}) $\sigma(p_{1}\,, p_{2})$ is  the BES dressing factor \cite{Arutyunov:2004vx, Beisert:2006ez}.
The $S$-matrices in (\ref{S-matrices}) satisfy the unitarity condition ($S_{AA}^{12}S_{AA}^{21}=\mathbb{I}$)  and the Yang-Baxter equations.
In particular, the scalar factors $S_{0}, \tilde{S}_{0}$  fulfil separately the unitary constraint and obey the following crossing relations:
\be
S_{0}(p_{1},p_{2})\tilde{S}_{0}({\bar{p}_{1},p_{2}})=S_{0}(p_{1},p_{2})\tilde{S}_{0}({p_{1},\bar{p}_{2}})=f(p_{1},p_{2})~,
\label{cross}
\ee
with
\be
f(p_{1},p_{2}) = \frac{\left(\frac{1}{x^{-}_{1}} -
x^{-}_{2}\right)(x^{-}_{1} - x^{+}_{2})}
{\left(\frac{1}{x^{+}_{1}} -
x^{-}_{2}\right)(x^{+}_{1} - x^{+}_{2})}~,
\ee
and $x^{\pm}(\bar{p})=1/x^{\pm}(p)$. Using the elliptic parametrisation equation (\ref{cross})   becomes
\be
S_{0}(z_{1},z_{2})\tilde{S}_{0}({z_{1}+\omega_{2},z_{2}})=S_{0}(z_{1},z_{2})\tilde{S}_{0}({z_{1},z_{2}-\omega_{2}})=f(z_{1},z_{2})~.
\ee
Since $S(p_1, p_2)$ coincides with the  $SU(2|2)$ $S$-matrix of $\text{AdS}_5/\text{CFT}_4$, one can repeat the analysis of \cite{Arutyunov:2007tc} concerning the underlying supersymmetry algebra, the parametrisation of this $S$-matrix and its mirror version.
The only difference occurs when we consider the scalar factor: as we can see in (\ref{cross}), the crossing relation does not relate a $S$-matrix with itself, but rather $S_{AA}$ ($S_{BB}$) with $S_{AB}$ ($S_{BA}$) and vice versa.
However, the $S$-matrices (\ref{S-matrices}) continue to satisfy the invariance under simultaneous shift of the arguments $z_1$ and $z_2$ by one half of the imaginary period:
\ba
S_{AA}(z_1+\omega_2, z_2+\omega_2)=S_{BB}(z_1, z_2)=S_{AA}(z_1, z_2)~,\nn\\
S_{AB}(z_1+\omega_2, z_2+\omega_2)=S_{BA}(z_1, z_2)=S_{AB}(z_1, z_2)~.\nn
\ea
Then we assume that the $S$-matrices (\ref{S-matrices}) share the good properties of
the $SU(2|2)$-related  scattering matrix  $S(p_1,p_2)$ and admit an analytic continuation in the $z$-variables, such that the mirror $S$-matrices are given by
\ba
\tilde{S}_{AA}(z_1, z_2)=\tilde{S}_{BB}(z_1, z_2)=S_{AA}(z_1+\frac{\omega_2}{2}, z_2+\frac{\omega_2}{2})=S_{BB}(z_1+\frac{\omega_2}{2}, z_2+\frac{\omega_2}{2})~,\nn\\
\tilde{S}_{AB}(z_1, z_2)=\tilde{S}_{BA}(z_1, z_2)=S_{AB}(z_1+\frac{\omega_2}{2}, z_2+\frac{\omega_2}{2})=S_{BA}(z_1+\frac{\omega_2}{2}, z_2+\frac{\omega_2}{2})~.\nn
\ea
They result still to satisfy the generalised unitarity condition
\be
\left[\tilde{S}_{AA}(z_1, z_2)\right]^\dagger=\tilde{S}_{AA}(z_2^*,z_1^*)~,\nn
\ee
and the dressing factor in the mirror theory has been derived in \cite{Arutyunov:2009kf}.
\resection{The  Bethe Ansatz equations}
\label{section4}
The asymptotic Bethe Ansatz equations for the $\mathcal{N}=6$  superconformal Chern-Simons theory are \cite{Gromov:2008qe}
\bea
e^{i p^{A}_{k}J}
&=&\prod_{l=1 \atop l \ne k}^{K^{\mathrm{I}}_{A}} \left(\frac{u_{k}^{A} - u_{l}^{A}+\frac{2i}{h}}
{u_{k}^{A} - u_{l}^{A}-\frac{2i}{h}} \right)\left(\frac{x_{k}^{A-}-x_{l}^{A+}}{x_{k}^{A+}-x_{l}^{A-}}\right)^{\frac{1-\eta}{2}}\left(\sqrt{\frac{x_{l}^{A+}}{x_{l}^{A-}}\frac{x_{k}^{A-}}{x_{k}^{A+}}}\right)^{\frac{1+\eta}{2}}\sigma(p^{A}_{k}\,, p^{A}_{l})\nonumber\\
&\times&
\prod_{l=1}^{K^{\mathrm{I}}_{B}} \left(\frac{x_{k}^{A-}-x_{l}^{B+}}{x_{k}^{A+}-x_{l}^{B-}}\right)^{\frac{1-\eta}{2}}\left(\sqrt{\frac{x_{l}^{B+}}{x_{l}^{B-}}\frac{x_{k}^{B-}}{x_{k}^{B+}}}\right)^{\frac{1+\eta}{2}}\sigma(p^{A}_{k}\,, p^{B}_{l})\,
\prod_{j=1}^{K^{\mathrm{II}}}
\left(\left(\frac{x_{k}^{A-}-y_j}
{x_{k}^{+A}-y_j} \right)\sqrt{\frac{x_{k}^{A+}}{x_{k}^{A-}}}\right)^{\eta}~,\nn\\
&&(k = 1, \ldots, K^{\mathrm{I}}_{A})
\eea
\bea
e^{i p^{B}_{k}J}
&=&\prod_{l=1 \atop l \ne k}^{K^{\mathrm{I}}_{B}} \left(\frac{u_{k}^{B} - u_{l}^{B}+\frac{2i}{h}}
{u_{k}^{B} - u_{l}^{B}-\frac{2i}{h}} \right)\left(\frac{x_{k}^{B-}-x_{l}^{B+}}{x_{k}^{B+}-x_{l}^{B-}}\right)^{\frac{1-\eta}{2}}\left(\sqrt{\frac{x_{l}^{B+}}{x_{l}^{B-}}\frac{x_{k}^{B-}}{x_{k}^{B+}}}\right)^{\frac{1+\eta}{2}}\sigma(p^{B}_{k}\,, p^{B}_{l}) \nonumber\\
& \times &
\prod_{l=1}^{K^{\mathrm{I}}_{A}}\left(\frac{x_{k}^{B-}-x_{l}^{A+}}{x_{k}^{B+}-x_{l}^{A-}}\right)^{\frac{1-\eta}{2}}\left(\sqrt{\frac{x_{l}^{A+}}{x_{l}^{A-}}\frac{x_{k}^{A-}}{x_{k}^{A+}}}\right)^{\frac{1+\eta}{2}} \sigma(p^{B}_{k}\,, p^{A}_{l})\,
\prod_{j=1}^{K^{\mathrm{II}}}
\left(\frac{x_{k}^{B-}-y_j}
{x_{k}^{B+}-y_j}\sqrt{\frac{x_{k}^{B+}}{x_{k}^{B-}}}\right)^{\eta}~,\nn\\
&&(k = 1, \ldots, K^{\mathrm{I}}_{B})\\
1&=&\prod_{l=1}^{K^{\mathrm{I}}_A} \left(\frac{y_{k} - x_{l}^{A+}}
{y_{k} - x_{l}^{A-}}\right)\sqrt{\frac{x_{l}^{A-}}{x_{l}^{A+}}}
\prod_{l=1}^{K^{\mathrm{I}}_{B}} \left(\frac{y_{k} - x_{l}^{B+}}
{y_{k} - x_{l}^{B-}}\right)\sqrt{\frac{x_{l}^{B-}}{x_{l}^{B+}}}
 \prod_{l=1}^{K^{\mathrm{III}}}
 \left( \frac{v_k
- w_{l} + \frac{i}{h}}
{v_k
- w_{l} - \frac{i}{h}}\right)~,\nn\\
&&(k = 1,\dots, K^{\mathrm{II}}),
\eea
\bea
1&=&
\prod_{l=1}^{K^{\mathrm{II}}} \left(\frac{w_{k} -
v_l-\frac{i}{h}}
{w_{k} -
v_l+\frac{i}{h}} \right)
\prod_{l=1 \atop l\ne k}^{K^{\mathrm{III}}}
\left(\frac{w_{k} - w_{l} + \frac{2i}{h}}
{w_{k} - w_{l} - \frac{2i}{h}}\right)~,~~~~ (k = 1, \dots,
K^{\mathrm{III}})~,
\label{BAE}
\eea
where
\eq
x_{k}^{\a\pm}=\frac{u^{\a}_{k}\pm\frac{i}{h}}{2}\left(1+\sqrt{1-\frac{4}{\left(u^{\a}_k\pm\frac{i}{h}\right)^{2}}}
\right)~,
\en
with $\a=A, B$, $h=h(\lambda)$ and $\eta=\pm1$ is the grading related to two different Dynkin diagrams.
For $\eta=1$ the momentum-carrying Bethe roots are bosonic, they can form single species bound states and the two massive nodes are linked  to each other only through a single BES dressing factor. In the $\eta=-1$ case, instead, the momentum-carrying roots are fermionic and cannot form bound states in the physical theory.

We see that the form of the asymptotic Bethe Ansatz equations crucially depends  on the grading $\eta$ or, equivalently, on the
choice of the reference state. In \cite{Arutyunov:2007tc} the $sl(2)$ grading $\eta=-1$ was chosen for the mirror BA equations. Similarly, we choose here the $sl(2|1)$ grading for the mirror BAEs:

\bea
\la{BY1}
e^{i\tp_{k}^{A} R} &=&\prod_ {l=1}
^{K_{A}^{\mathrm{I}}}S_0(\tp_{k}^{A},\tp_{l}^{A})\prod_ {l=1}^{K_{B}^{\mathrm{I}}}\tilde{S}_{0}(\tp_{k}^{A},\tp_{l}^{B})
\prod_{l=1}^{K^{\mathrm{II}}}\left(\frac{x_{k}^{A+}-y_l}
{x_{k}^{-A}-y_l}\right)\sqrt{\frac{x_{k}^{A-}}{x_{k}^{A+}}}~,\\
\la{BY2}
e^{i\tp_{k}^{B} R} &=&\prod_ {l=1}
^{K_{B}^{\mathrm{I}}}S_0(\tp_{k}^{B},\tp_{l}^{B})\prod_ {l=1}^{K_{A}^{\mathrm{I}}}\tilde{S}_{0}(\tp_{k}^{B},\tp_{l}^{A})
\prod_{l=1}^{K^{\mathrm{II}}}\left(\frac{x_{k}^{B+}-y_l}
{x_{k}^{B-}-y_l}\right)\sqrt{\frac{x_{k}^{B-}}{x_{k}^{B+}}}~,\\
\label{BYfm1}
-1&=&\prod_{\a=A,B}\prod_{l=1}^{K_{\a}^{\mathrm{I}}} \left(\frac{y_{k}-x^{\a+}_{l}}{y_{k}-x^{\a-}_{l}} \right)\sqrt{\frac{x_l^{\a-}}{x_l^{\a+}}}
\prod_{l=1}^{K^{\mathrm{III}}}\left(\frac{v_{k}-w_{l}+\frac{i}{h}}{v_{k}-w_{l}-\frac{i}{h}}\right)~,\\
1&=&\prod_{l=1}^{K^{\mathrm{II}}}\left(\frac{w_{k}-v_{l}-\frac{i}{h}}{w_{k}-v_{l}+\frac{i}{h}}
\right) \prod_ {\textstyle\atopfrac{l=1}{l\neq
k}}^{K^{\mathrm{III}}} \left(\frac{w_{k}-w_{l}+\frac{2i}{h}}{w_{k}-w_{l}-\frac{2i}{h}} \right)~,
\label{BYf}
\eea
where
\ba
S_0(\tp_{k}^{\a},\tp_{l}^{\a})&=& \left(
\frac{u^{\a}_{k}-u^{\a}_{l}+\frac{2i}{h}}{u^{\a}_{k}-u^{\a}_{l}-\frac{2i}{h}}
\right)\, \left(\frac{x_k^{\a-}-x_l^{\a+}}{x_k^{\a+}-x_l^{\a-}}\right)\,\sigma(\tp_{k}^{\a},\tp_{l}^{\a})~,\\
\tilde{S}_0(\tp_{k}^{\a},\tp_{l}^{\beta})&=&
\left(\frac{x_k^{\a-}-x_l^{\beta+}}{x_k^{\a+}-x_l^{\beta-}} \right)\,\sigma(\tp_{k}^{\a},\tp_{l}^{\beta})~.
\ea
Notice that for the  y-particles,
$
y_k= i e^{- i q_k}
$
and exchanging   $q_k \leftrightarrow \pi-q_k$~$\hbox{mod}(2 \pi)$ corresponds to
$
y_k \leftrightarrow 1/y_k.
$
Considering the latter property, and in order to write  the final  equations in a more transparent form,  it is convenient   to define
the multi-valued function $y(u)$ such that
\eq
y(u) =  \left\{ \begin{array}{lll}
 x(u) & \hbox{for} & \Im m(y)<0~,  \\
& & \\
1/x(u) & \hbox{for} & \Im m(y)>0~,
\end{array}\right.
\label{yy}
\en
where   $x(u)$ is defined by
\eq
x(u)= \left( \frac{u}{2} -i \sqrt{ 1- \frac{u^2}{4}} \right)~,
\en
and  the  property $x(u)=1/x^*(u^*)$. Setting $u(q)=2\sin(q)$ and
\eq
u^{y|+}_k=u(q_k)~,
\en
with  $q_k \in (-\pi/2,\pi/2]$, or equivalently
$\Im m(y_k)> 0$, we have
\eq
y_k=y(u^{y|+}_k) \equiv 1/x(u(q_k))~.
\en
Similarly  for $q_k \in (-\pi,-\pi/2] \cup (\pi/2,\pi]$, or equivalently $\Im m(y_k)< 0$:
\eq
y_k=y(u^{y|-}_k)= y((u(q_k) +2)e^{i 2 \pi}-2)) \equiv x(u(q_k))~.
\en

\resection{The string hypothesis}
\label{section5}
Considering  the pole structure of equations (\ref{BY1}-\ref{BYf})  and  the mathematical resemblance between
equations (\ref{BYfm1}) and (\ref{BYf}) with those of an inhomogeneous Hubbard model, we can formulate a string hypothesis for the solutions, in strict analogy with the Takahashi' s one~\cite{Takahashi}. We shall assume that the thermodynamically relevant solutions
of (\ref{BY1}-\ref{BYf}) in the limit of large $R, K^I_{\a}, K^{II}, K^{III}$ rearrange themselves into complexes -- the so-called strings -- with real centres and all the other
complex roots symmetrically distributed around these centres along the imaginary direction.

In particular, the strings of the `massive' Bethe roots are  those of an $sl(2|1)$ spin chain (see \cite{Saleur:2009bf} for a discussion on the string hypothesis and the TBA for this model), with mixed, alternating configurations of roots of kind $A$ and $B$.
In fact, they arise from considering the following parts of the equations (\ref{BY1}) and (\ref{BY2}), where we set all the auxiliary roots to zero:

\ba
e^{i\tp_{k}^{A} R} &=&\prod_ {l=1}
^{K_{A}^{\mathrm{I}}}\frac{1-\frac{1}{x_k^{A+}x_l^{A-}}}{1-\frac{1}{x_k^{A-}x_l^{A+}}}\,\sigma(\tp_{k}^{A},\tp_{l}^{A})\prod_ {l=1}^{K_{B}^{\mathrm{I}}}\frac{x_k^{A-}-x_l^{B+}}{x_k^{A+}-x_l^{B-}}\,\sigma(\tp_{k}^{A},\tp_{l}^{B})~,\\
e^{i\tp_{k}^{B} R} &=&\prod_ {l=1}
^{K_{B}^{\mathrm{I}}}\frac{1-\frac{1}{x_k^{B+}x_l^{B-}}}{1-\frac{1}{x_k^{B-}x_l^{B+}}}\,\sigma(\tp_{k}^{B},\tp_{l}^{B})\prod_ {l=1}^{K_{A}^{\mathrm{I}}}\frac{x_k^{B-}-x_l^{A+}}{x_k^{B+}-x_l^{A-}}\,\sigma(\tp_{k}^{B},\tp_{l}^{A})~.
\ea
When $R\rightarrow\infty$, the following two-particle bound state equations hold (let us suppress the indexes $k=l$ for simplicity)
\be
x^{A+}-x^{B-}=0\quad\mbox{or}\quad x_1^{B+}-x_1^{A-}=0~.
\ee
The first equation is solved by $\tilde{p}^A=p/2-iq, \tilde{p}^B=p/2+iq$, with $\Re e(q)>0$, or $u^B-u^A=2i/h$, the second one by $u^A-u^B=2i/h$.
We verified that the possible bound states equations arising from the other denominators, of the type $1-\frac{1}{x_k^{\a-}x_l^{\a+}}=0$, do not give `good solutions' and assume that also the poles of the dressing factor do not give physical bound states.

Thus, thanks to the great similarity of the remaining part of the equations (\ref{BY1}) with the BA equations of the $\text{AdS}_5\times S^5$ mirror model \cite{Arutyunov:2007tc, Arutyunov:2009zu} and considering  that
\eq
\tilde{p}^Q_k=\tilde{p}^Q(u_k) \equiv \tilde{p}(u_k) = \frac{i h}{2} \left( \sqrt{4 - \left(u_k+ i \fract{Q}{h}\right)^2}
-\sqrt{4 - \left(u_k- i \fract{Q}{h}\right)^2} \right)~, \label{pt}\\
\en
we can propose the following string classification:
\begin{itemize}
\item Wide strings\\
I) $N_{Q|{WI}}$ $WI$-particles (bound states) with real momenta $\tilde{p}^{Q|{WI}}_k=\tilde{p}_k^{2Q-1}$ and real rapidities $u_{k}^{{Q|WI}}$:
\bea
\label{strings1}
u_{k,j}^{A\,Q}&=&u_k^{{Q|WI}}+\frac{i}{h}\,(2Q+2-4j)~,~~~j=1,\dots,Q\ ;\\
u_{k,l}^{B\,Q}&=&u_k^{{Q|WI}}+\frac{i}{h}\,(2Q-4l)~,~~~~~~~~~l=1,\dots,Q-1\ .
\eea
II) $N_{Q|{WII}}$ ${WII}$-particles (bound states) with real momenta $\tilde{p}^{Q|{WII}}_k=\tilde{p}_k^{2Q-1}$ and real rapidities $u_{k}^{{Q|WII}}$:
\bea
u_{k,j}^{A\,Q}&=&u_k^{{Q|WII}}+\frac{i}{h}\,(2Q-4j)~,~~~~~~~~j=1,\dots,Q-1\ ;\\
u_{k,l}^{B\,Q}&=&u_k^{{Q|WII}}+\frac{i}{h}\,(2Q+2-4l)~,~~~~l=1,\dots,Q\ .
\eea
\item Strange strings\\
I) $N_{Q|{sI}}$ ${sI}$-particles (bound states) with real momenta $\tilde{p}^{Q|{sI}}_k=\tilde{p}_k^{2Q}$ and real rapidities $u_{k}^{{Q|sI}}$:
\bea
u_{k,j}^{A\,Q}&=&u_k^{{Q|sI}}-\frac{i}{h}\,(2Q+1-4j)~,~~~~j=1,\dots,Q\ ;\\
u_{k,l}^{B\,Q}&=&u_k^{{Q|sI}}+\frac{i}{h}\,(2Q+ 1-4l)~,~~~~~l=1,\dots,Q\ .
\eea
II) $N_{Q|{sII}}$ ${sII}$-particles (bound states) with real momenta $\tilde{p}^{Q|{sII}}_k$ and real rapidities $u_{k}^{{Q|sII}}$:
\bea
u_{k,j}^{A\,Q} &=& u_k^{{Q|sII}}+\frac{i}{h}\,(2Q+1-4j)~,~~~~j=1,\dots,Q\ ;\\
u_{k,l}^{B\,Q} &=&u_k^{{Q|sII}}-\frac{i}{h}\,(2Q+1-4l)~,~~~~~ l=1,\dots,Q\ .
\eea
\item $N_y$ $y$-particles with real momenta $q_k \in (-\pi,\pi]$.\\
\item $N_{M|v}$ $vw$-strings with real centers $v_k^M$, $2M$ roots of type $v$ and M of type $w$:
\bea
v_k^{M,j}&=&v_k^M\pm\frac{i}{h}(M+2-2j)~,~~~~~~j=1,\dots,M\ ;\\
w_k^{M,j}&=&v_k^M+\frac{i}{h}(M+1-2j)~,~~~~~~j=1,\dots,M\ .
\eea
\item $N_{N|w}$ $w$-strings with real centres $w_k^N$ and $N$ roots of type $w$:
\bea
\label{strings2}
w_k^{N,j}&=&w_k^N+\frac{i}{h}(N+1-2j)~,~~~~j=1,\dots,N\ .
\eea
\end{itemize}
Notice that the strings $WI$ have Bethe roots of type A as real centres, the $WII$ ones have Bethe roots of type B as real centres; the strange strings have not real centres belonging to the Bethe roots and are not invariant under complex conjugation.
We will return on this peculiar property in the next section, when we will discuss the root density equations.
Replacing the  variables $u^{A,B}_k, v_k$ and $w_k$ in (\ref{BY1})  with $u_{k,j}^{Q}, v_k^{M,j}$ and $w_k^{M,j}$, performing the products on the internal string index $j$, and
relabelling   the Q-related quantities as
\eq
{ \CQ|\a}  =  \left\{ \begin{array}{lll}
 { Q|W \a} & \hbox{for} & \CQ~= 2Q-1~; \\
& & \\
{ Q|s \a} & \hbox{for} & \CQ~= 2Q~,
\end{array}\right.~
\en
we arrive  to the following equations for the real centres of the  strings (\ref{strings1}-\ref{strings2}):

\ba
1&=&e^{i\tilde{p}_{k}^{\CQ|\a}R}\prod_{\beta,\CQ'}\prod_{l=1}
^{N_{\CQ'|{\beta}}}S_{(\CQ|\a),(\CQ'|\beta)}(u_{k}^{\CQ|\a} ,u_{l}^{\CQ'|\beta})
\prod_{M} \prod_{l=1}^{N_{M|vw}} S_{\CQ,(v|M)} (u_{k}^{\CQ|\a},v_{l}^M) \nn \\
&\times& \prod_{\delta=\pm}\prod_{l=1}^{N_{y|\delta}} S_{\CQ,y}(u_{k}^{\CQ|\a},u^{y|\delta}_{l})
\,,\label{stringWI}\\
-1&=&\prod_{\beta,\CQ}\prod_{l=1}^{N_{\CQ|{\beta}}}S_{y,\CQ}(u^{y|\pm}_{k},u_{l}^{\CQ|\beta})
\prod_{M} \prod_{l=1}^{N_{M|w}}S_{M}(u^{y|\pm}_{k}-w_{l}^M) \,
 \prod_{l=1}^{N_{M|vw}}  S_{M}(u^{y|\pm}_{k}-v_{l}^M) \,,
\label{string2}\\
(-1)^{K}&=&\prod_{\beta,\CQ}\prod_{l=1}^{N_{\CQ|\beta}}S_{(v|K),\CQ}(v_{k}^K,u_{l}^{\CQ|\beta}) \prod_{M} \prod_{l=1}^{N_{M|vw}}S_{KM}(v_{k}^K-v_{l}^M) \prod_{\delta=\pm}\prod_{l=1}^{N_{y|\delta}} S_{K}(v_{k}^K- u^{y|\delta}_{l})\,,~~~~~~~\label{string3}\\
  (-1)^K&=& \prod_{M} \prod_{l=1}^{N_{w|M}}S_{KM}(w^K_{k}-w^M_{l})\prod_{\delta=\pm}\prod_{l=1}^{N_{y|\delta}} \left(S_{K}(w^K_{k}-u^{y|\delta}_l)) \right)^{-1}
 \,,~~~~~\label{string4}
\ea
with $\CQ=1,2,\dots$ and $\alpha=I,II$. The scalar $S_{AB}(u,z)$ factors in  (\ref{stringWI}--\ref{string4}) are
listed in  Appendix A.

\resection{The thermodynamic Bethe Ansatz method}
\label{section6}
The mirror Bethe Ansatz equations (\ref{stringWI}-\ref{string4}) can be used to derive,  using
a procedure \cite{YY, Takahashi, onedHubbard, Zamolodchikov:1989cf}
 already successfully adapted to the study of  $\text{AdS}_5/\text{CFT}_4$ in \cite{Bombardelli:2009ns,Gromov:2009bc, Arutyunov:2009ur}, a set of thermodynamic Bethe Ansatz equations describing the ground-state of the (direct) $\text{AdS}_4/\text{CFT}_3$ theory. Taking the logarithm of  (\ref{stringWI}-\ref{string4}), and introducing  the collective index $A$  for the different density labels we can perform  the thermodynamic limit $N_A, R\rightarrow\infty$  with $N_A/ R$ finite. The density of states $\rho_A$  is
\eq
\rho_A(u)=\rho^r_A(u)+\rho^h_A(u)=\lim_{R\rightarrow\infty}\frac{I^{A}_{k+1}-I^{A}_k}{R (u^{A}_{k+1}-u^A_{k})}~,
\label{rho}
\en
where $\rho^r_A$ and $\rho^h_A$ are respectively the density of roots and holes
  and the  $I$'s are the Bethe  quantum numbers.  $I_k \in \ZZ$ for $A \in \{(\CQ|\a), (y|\pm) \}$ with $\alpha=I,II$ and $\CQ=1,2,\dots$ while  $I_k \in \ZZ+1/2$   for  $A \in \{(v|K),(w|K) \} $ with $K=1,2,\dots$.

The Bethe Ansatz equations (\ref{stringWI}--\ref{string4}) lead to a set of constrains for the  densities (\ref{rho}):
\ba
\label{dens}
\rho_{\CQ|{\a}}(u)&=&\frac{1}{2\pi} {d \tilde{p}^{\CQ}(u) \over du}+\sum_{\beta}\sum_{\CQ'=1}^{\infty}
\phi_{(\CQ|\a),(\CQ'|\beta)}*\rho_{\CQ'|{\beta}}^r(u)+\sum_{M=1}^{\infty}\phi_{\CQ,(v|M)}*\rho_{v|M}^{r}(u) \nn \\
&+& \int_{-2}^{2} dz\left[\phi_{\CQ,(y|-)}(u,z)\rho_{y|-}^{r}(z)+\phi_{\CQ,(y|+)}(u,z)\rho_{y|+}^{r}(z)\right]~, \\
\rho_{y|-}(u)&=&-\sum_{\CQ=1}^{\infty}\phi_{(y|-),\CQ}*(\rho_{\CQ|I}^{r}(u)+\rho_{\CQ|II}^{r}(u))
-\sum_{M=1}^{\infty}\phi_{M}*(\rho_{w|M}^{r}(u)+ \rho_{v|M}^{r}(u))\ ,\\
\rho_{v|K}(u)&=&-\sum_{\CQ=1}^{\infty} \phi_{(v|K),\CQ}*(\rho_{\CQ|I}^{r}(u)+\rho_{\CQ|II}^{r}(u) )-\sum_{M=1}^{\infty} \phi_{K,M}*\rho_{v|M}^{r}(u) \nn \\
&-& \int_{-2}^{2}dz\, \phi_{K}(u-z)\, (\rho_{y|-}^{r}(z)+\rho_{y|+}^{r}(z))\ ,\\
\rho_{w|K}(u)&=&-\sum_{M=1}^{\infty}\phi_{K,M}*\rho_{w|M}^{r}(u)
+ \int_{-2}^{2}dz\, \phi_{K}(u-z)\, (\rho_{y|-}^{r}(z)+\rho_{y|+}^{r}(z))\ ,
\label{dens2}
\ea
where
\eq
\rho_{y|+}(u)= \rho_{y|-}((u+2) e^{i 2 \pi}-2)~,~~~
\rho^r_{y|+}(u)= \rho^r_{y|-}((u+2) e^{i 2 \pi}-2)~,
\en
and the symbol `$*$' denotes the convolution
\eq
\phi*\rho(u)=\int_{\RR} dz \,\phi(u,z)\,\rho(z)~.
\en
The kernels are
\eq
\phi_{AB}(z,u)= {1 \over 2 \pi i} \frac{ \partial}{\partial z} \ln S_{AB}(z,u)~.
\en
 Finally, as anticipated in the previous section, the strange strings are not invariant under complex conjugation
 and  to define real  densities $\rho_{2Q|\a}(u)$, we have to impose the equality between the real centres of the strange strings of type I and type II. This corresponds to
\eq
\rho_{2Q|I}(u)=\rho_{2Q|II}(u)~.
\en
\subsection{The TBA equations}
\label{TBAe}
In terms of  hole and root densities,  the entropy is
\eq
S =\sum_{A} \int du \; \left((\rho^r_{A}(u)+\rho^h_{A}(u))\ln(\rho^r_{A}(u) +\rho^h_{A}(u))
-\rho^r_{A}(u)\ln\rho^r_{A}(u) \right)~,
\en
and  the free energy per unit length:
\be
f(T)=\tilde{H}-TS~.
\label{rhofree}
\ee
In (\ref{rhofree}) $\tilde{H}$ is the (mirror) energy per unit length:
\eq
\tilde{H}= \sum_{\CQ=1}^{\infty}\int_{\RR} du \; E_{\CQ}(u) (\rho_{\CQ|I}^r(u)+\rho_{\CQ|II}^r(u))~,
\en
with
\eq
E_{\CQ}(u) = \ln{\frac{x(u-\frac{i}{h}\CQ)}{x(u+\frac{i}{h}\CQ )}}~.
\en
The temperature $T$ of the mirror theory corresponds to the  inverse of the trace operator length  $L$ in
${\cal N}=6$ superconformal Chern-Simons theory. The extremum condition $\delta f=0$ under the constraints (\ref{dens}-\ref{dens2}) leads  to the set of TBA equations for the pseudoenergies $\vep_A(u)$:
\eq
\vep_A(u)=\ln {\rho^h_A(u) \over \rho^r_A(u)}~,~~~{1 \over e^{\vep_A(u)}+1}= {\rho^r_A(u) \over \rho_A(u)}~,~~~
L_A(u)=\ln\left(1+e^{-\vep_A(u)} \right)~.
\label{dfde}
\en
The  TBA  equations are
\ba
\vep_{\CQ|{\a}}(u)&=&L\,E_{\CQ}(u)
-\sum_{\beta}\sum_{\CQ'=1}^{\infty}L_{\CQ'|\beta}*\phi_{(\CQ'|\beta),(\CQ|\a)}(u)
+\sum_{M=1}^{\infty}L_{v|M}*\phi_{(v|M),\CQ}(u) \nn \\
&+&
\int_{-2}^{2} dz\left[L_{y|-}(z)\,\phi_{(y|-),\CQ}(z,u)
- L_{y|+}(z)\,\phi_{(y|+),\CQ} (z,u)\right]
~,\label{TBA1a} \\
\vep_{y|-}(u)&=&-\sum_{\CQ=1}^{\infty}(L_{\CQ|I }+L_{\CQ|II})*\phi_{\CQ,(y|-)}(u)
+ \sum_{M=1}^{\infty} (L_{v|N}-L_{w|M}  )*\phi_{M}(u)~, \label{TBA2}\\
\vep_{v|K}(u)&=&-
\sum_{\CQ=1}^{\infty}(L_{Q|I}+L_{Q|II})*\phi_{\CQ,(v|K)}(u)
+\sum_{M=1}^{\infty}L_{v|M}*\phi_{M,K}(u) \nn \\
&+& \int_{-2}^{2} dz\,(L_{y|-}(z)-L_{y|+}(z))\, \phi_K(z-u)
~,~~~\label{TBA3} \\
\vep_{w|K}(u)&=&\sum_{M=1}^{\infty}L_{w|M}*\phi_{M,K}(u)+ \int_{-2}^{2} dz\,(L_{y|-}(z)-L_{y|+}(z))\, \phi_{K}(z-u)
~,
\label{TBA4}
\ea
where
\eq
\vep_{y|+}(u)=  \vep_{y|-}((u+2) e^{i 2 \pi}-2)~,
\en
and
\eq
L*\phi(u)= \int_{\RR} dz \, L(z) \phi(z,u)~.
\en
Finally, the {\it minimal} free energy  is given by the following non-linear functional of the pseudoenergies $\vep_{\CQ|{\a}}(u)$
\ba
f(T)=-T\sum_{\CQ=1}^{\infty}\int_{\RR} {du \over 2 \pi}  \,  {d\tilde{p}^{\CQ} \over du}  ( L_{\CQ|I}(u)+L_{\CQ|II}(u))~,
\ea
where $f(T)$ is related  to  the ground state energy for the $\text{AdS}/\text{CFT}$ theory on a circumference with length $L=1/T$  by the relation
\eq
E_0(L)=Lf(1/L)~.
\en
As we have kept the total densities finite, it is natural to introduce chemical potentials $\mu_A$. For  relativistic theories this has been  discussed in~\cite{Klassen:1990dx}.
The TBA equations (\ref{TBA1a}--\ref{TBA4}) do not change their form, but for this simple replacement
\eq
L_A=\ln (1+ e^{-\epsilon_A})  \rightarrow L_{A,\lambda}=\ln (1+ \lambda_A e^{-\epsilon_A})~,
\label{Ldef}
\en
involving the fugacities $\lambda_A=e^{\mu_A/T}$.

In our case we expect zero energy as soon as the fugacities reach these values
\eq
\lambda_{\CQ|\a}=(-1)^\CQ ~,~~\lambda_{v|K}=\lambda_{w|K}=1~,
~~\lambda_{y|\pm}=-1~,~~~(\a =I,II,~K=1,2,\dots)~.
\label{fugacities}
\en

Physically, this modification corresponds to the computation of the Witten
index and the proposal  (\ref{fugacities}) reflects  the  bosonic/fermionic  character
of    the various excitations. A vanishing ground state energy can be given by a singularity in
the solution of the massive TBA equation (\ref{TBA1a}). Of course, the latter needs to be 
regularised by introducing the chemical potentials
\eq
\lambda_{2Q-1|I}= -e^{ih}~,~~\lambda_{2Q-1|II}= - e^{-ih}~,~~\lambda_{2Q|I}= \lambda_{2Q|II}= 1~,~~\lambda_{v|K}=\lambda_{w|K}=1~,~~\lambda_{y|\pm}=-1,
\label{fugacities1}
\en
such that the TBA equations are regular for $h \ne 0$ and the ground
state energy tends to zero as  $h\rightarrow 0$. It would be important to check 
directly the vanishing of $E_0(L)$ using numerical or analytic methods.

\section{The Y-system}
\label{section7}

Using the methods adopted in \cite{Ravanini:1992fi, Bombardelli:2009ns, Arutyunov:2009ur} and  the kernel identities listed in  Appendix A, the  following
Y-system valid in the strip $|\Re e(u)| < 2$ can be derived. For the $WI-$, $WII-$, $sI-$ and $sII-$related
pseudoenergies we have
\bea
Y_{1|I}(u+\fract{i}{h})Y_{1|II}(u-\fract{i}{h}) &=& (1+Y_{2|I}(u))
\left( 1 + {1 \over Y_{y|-}(u)} \right)^{-1}~,
\label{Ys6}\\
Y_{1|II}(u+\fract{i}{h})Y_{1|I}(u-\fract{i}{h}) &=& (1+Y_{2|II}(u))
\left( 1 + {1 \over Y_{y|-}(u)} \right)^{-1}~,
\eea
\ba
Y_{\CQ|I}(u+\fract{i}{h})Y_{\CQ|II}(u-\fract{i}{h}) &=& (1+Y_{\CQ+1|I}(u)) (1+Y_{\CQ-1|II}(u))
\left( 1 + {1 \over Y_{v|\CQ-1}(u)} \right)^{-1}~,~~~~~
\label{Ys5}\\
Y_{\CQ|II}(u+\fract{i}{h})Y_{\CQ|I}(u-\fract{i}{h}) &=& (1+Y_{\CQ+1|II}(u)) (1+Y_{\CQ-1|I}(u))
\left( 1 + {1 \over Y_{v|\CQ-1}(u)} \right)^{-1}~,~~~~~
\label{Ys9}
\ea
with $Y_A=e^{\ep_A}$ and $\CQ=2,3,\dots$. The reader should notice that  equations (\ref{Ys6}-\ref{Ys9}) have a slightly different
structure compared to the standard Y-systems as, for example, those proposed for the same model in \cite{GKVI}. The right-hand-sides of (\ref{Ys6}-\ref{Ys9}) involve mixed pairs of type $I$ and type $II$  functions. A similar non-standard structure has been previously observed in the context of a much-simpler  $D_n$-related family
of Y-systems~\cite{Caracciolo:1999ih}. In the current case we  suspect this property should be  directly related to the  multi-valued character  of the Y-functions. The equations for the remaining TBA nodes are:

\bea
Y_{w|K}(u+\fract{i}{h})Y_{w|K}(u-\fract{i}{h}) &=& \prod_{M=1}^{\infty}\left(1+{Y_{w|M}(u)} \right)^{I_{KM}} \left({ 1 + \frac{1}{Y_{y|-}(u)} \over 1 + \frac{1}{Y_{y|+}(u)} } \right)^{\delta_{K1}}~,
\label{Ys1} \\
Y_{y|-}(u+\fract{i}{h}) Y_{y|-}(u-\fract{i}{h}) &=& { \left(1+Y_{v|1}(u) \over
 1 + Y_{w|1}(u) \right)}
\prod_{\a=I,II}\left( 1 + {1 \over Y_{1|\a}(u)} \right)^{-1}~,
\label{Ys4} \\
Y_{v|K}(u+\fract{i}{h})Y_{v|K}(u-\fract{i}{h}) &=& {\prod_{M=1}^{\infty}\left(1+{Y_{v|M}(u)} \right)^{I_{KM}}
 \over \prod_{\a} \left( 1+ \frac{1}{Y_{K+1|\a}}(u) \right)}
 \left({ 1 + Y_{y|-}(u) \over 1 + Y_{y|+}(u) } \right)^{\delta_{K1}}~,
\label{Yk}
\eea
with $K=1,2,\dots$. Although equations   (\ref{Ys6}--\ref{Yk}) were derived restricting  $u$ to the region  $|\Re e(u)| < 2$, they are obviously  valid in a  much wider region of the complex plane.  However  due to the presence of an infinite number of square-root branch points, the Y-functions are multi-valued and the analytic continuation of (\ref{Ys6}--\ref{Yk})
outside the region  $|\Re e(u)| < 2$   requires special attention~\cite{Arutyunov:2009ur}. A more complete discussion of the analytic properties of the solutions of the TBA equations (\ref{TBA1a}--\ref{TBA4}) is   postponed to the near future.

\resection{Conclusions}
\label{section8}

The study of the anomalous dimensions of single trace composite operators in the planar $\mathcal{N}=6$ superconformal Chern-Simons gauge theory in three dimensions is a very challenging objective.
In this paper, following some analogies with the more studied  ${\cal N}=4$ super Yang-Mills example, we have  proposed a set of all-loop Bethe Ansatz equations for the SCS {\it mirror} theory and formulated the corresponding string hypothesis. By means of these two ingredients it has been possible to derive a set of thermodynamic Bethe Ansatz equations for the ground-state energy and then an associated Y-system. Of course, the latter is less informative, and just because of its more generality it is conjectured to encompass excited state TBA, too.  

The $Y$-system here clearly differs from that of \cite{GKVI} in the momentum carrying nodes.
Actually, a $Y$-system like that proposed in \cite{GKVI}  can be derived by applying the TBA
procedure to the  {\it direct} $\text{AdS}_4\times\mathbb{CP}^3$ string theory. The latter is asymptotically described by the Bethe-Yang equations (\ref{BAE}) in the grading $\eta=1$, which imply $su(2)$-like bound states for the A particles and, separately, for the B ones. Therefore, the string hypothesis would not be different from that in \cite{Arutyunov:2009zu}, but for the presence of two species of particles, $A$ and $B$. Then, the Yang and Yang TBA procedure can be applied and the resulting TBA equations lead to a $Y$-system like that in \cite{GKVI}. In this way, this $Y$-system is deeply related to ours above\footnote{Moreover, in the $\text{AdS}_5/\text{CFT}_4$ correspondence this difference in the functional form of the $Y$-system (between direct and mirror theory) does not subsist. We are not referring here to the other differences, like, for instance, the integration domains in the TBA equations.}. Nevertheless, the two $Y$-systems share the same form as long as the massive nodes are identified, namely $Y_{\CQ|I}(u)=Y_{\CQ|II}(u)$ for the mirror case and  $Y_{Q|A}(u)=Y_{Q|B}(u)$ in the direct case, where $Y_{Q|A}(u)$ and $Y_{Q|B}(u)$ are the $Y$s of the $A$- and $B$-bound states, respectively\footnote{These are called respectively $Y_{a,0}^{4}(u)$ and $Y^{\bar{4}}_{a,0}(u)$ in \cite{GKVI}.}. Pictorially, this procedure folds the two massive node chains into a unique one. Provided that the asymptotic expressions \cite{GKVI} for $Y_{Q|A}(u)$ and $Y_{Q|B}(u)$ of the irrepresentation {\bf 20} are part of the $Y$-system solution, they should be solution of the mirror $Y$-system as well.

In the framework of relativistic scattering models a number of tools have been developed over the years~\cite{Bazhanov:1996aq, Dorey:1996re, Fioravanti:1996rz} to extend the equations from the ground state to the excited states. To verify the correctness and consistency of our proposals and  to make the connection with the field-theory results the generalisation to  excited states  is almost  compulsory. However, to achieve this objective, the analytic  properties of the  $Y$-functions should be understood at a deeper level. In particular the r\^ole of the  dressing factor should be clarified.  Some important progress in this direction have been recently made in \cite{Arutyunov:2009ax}, but we suspect  that the $\text{AdS}/\text{CFT}$-related TBA equations contain many more interesting surprises.

\section*{Acknowledgements}
We would like to thank Gleb Arutyunov, Andrea Cavagli\`a, Sergey Frolov and  Marco Rossi  for useful discussions and observations. We acknowledge the following grants: INFN {\it Iniziative specifiche FI11}, {\it PI11} and {\it PI14}, the international agreement INFN-MEC-2008 and the italian University PRIN 2007JHLPEZ ``Fisica Statistica dei Sistemi Fortemente Correlati all'Equilibrio e Fuori Equilibrio: Risultati Esatti e Metodi di Teoria dei Campi" for travel financial support.

\resection{Appendix~A}
%
Here we report  the scalar factors $S_{A,B}(u,z)$ involved in the Bethe Ansatz equations (\ref{stringWI}--\ref{string4}).
\bea
S_{y, \CQ}(u,z)&=&S_{\CQ, y}(z,u) =  \left(\frac{x(z-\frac{i}{h} \CQ)-y(u)}{x(z+\frac{i}{h}\CQ)-y(u)} \right) \sqrt{\frac{x(z+\frac{i}{h}\CQ)}{x(z-\frac{i}{h}\CQ)}} ~.
\label{SyQ}
\eea
The functions $S_{(y|\pm), \CQ}$ and  $S_{\CQ, (y|\pm)}$ are related to  (\ref{SyQ}) through equation (\ref{yy}), they are:
\eq
S_{(y|\mp), \CQ}(u,z)=S_{\CQ, (y|\mp)}(z,u) =  \left(\frac{x(z-\frac{i}{h} \CQ)-(x(u))^{\pm 1}}{x(z+\frac{i}{h}\CQ)-(x(u))^{\pm 1}}\right) \sqrt{\frac{x(z+\frac{i}{h}\CQ)}{x(z-\frac{i}{h}\CQ)}}~.
\en
Moreover, we may write down
\bea
S_{(v|M), \CQ}(u,z)&=&S_{\CQ,(v|M)}(z,u)= \left(
\frac{x(z-\frac{i}{h}\CQ)-x(u+\frac{i}{h}M )}{x(z+\frac{i}{h}\CQ)-x(u+ \frac{i}{h} M )}\right)
\left(\frac{x(z+\frac{i}{h}\CQ)}{x(z-\frac{i}{h}\CQ)} \right) \nn\\
&\times& \left(\frac{x(z-\frac{i}{h}\CQ)-x(u-\frac{i}{h} M)}{x(z+\frac{i}{h}\CQ)-x(u-\frac{i}{h}M )}\right)\prod_{j=1}^{M-1} \left(\frac{z-u-\frac{i}{h}(\CQ-M+2j)}{z-u+\frac{i}{h}(\CQ-M+2j)}\right)~,\\
S_M(u) &=&  \left(\frac{u-\frac{i}{h} M }{u+\frac{i}{h} M }
\right)~,
\eea
\eq
S_{K,M}(u)= \left( {u - \fract{i}{h} |K-M| \over u +\frac{i}{h} |K-M|} \right) \left(
 {u - \frac{i}{h} (K+M) \over u +\frac{i}{h}(K+M)} \right)
\prod_{k=1}^{\text{min}(K,M)-1} \left( {u - \frac{i}{h} (|K-M|+2k) \over u +\frac{i}{h}(|K-M|+2k)} \right)^2
~.~~~~
\en
The  elements $S_{(\CQ|\a),(\CQ'|\beta)}(u ,z)$ are:
\eq
S_{(\CQ|\alpha),(\CQ'|\beta)}(u,z)= S^0_{(\CQ|\alpha),(\CQ'|\beta)}(u-z) (\Sigma^{\CQ,\CQ'}(u,z))^{-1}~,
\en
where  $\Sigma^{\CQ,\CQ}$ is the improved dressing factor for the mirror bound states defined and derived in \cite{Arutyunov:2009kf}:
\eq
\Sigma^{\CQ,\CQ'}(u,z)=\prod_{k=1}^{\CQ}\prod_{l=1}^{\CQ'}\left(\frac{1-\frac{1}{x(u+\frac{i}{h}(\CQ+2-2k))x(z+\frac{i}{h}(\CQ'-2l))}}{1-\frac{1}{x(u+\frac{i}{h}(\CQ-2k))
x(z+\frac{i}{h}(\CQ'+2-2l))}}\right)\sigma^{\CQ, \CQ'}(u,z)~.
\en
Finally,  for $\alpha \ne \beta$ ($\alpha' \ne \beta', \alpha'' \ne \beta'', \alpha''' \ne \beta'''$) we have:
\bea
S^0_{(2Q-1|\a),(2Q'-1|\a)}(u) &=&\left(\frac{u+\frac{2i}{h}|Q'-Q|}{u-\frac{2i}{h}|Q'-Q|} \right)\prod_{j=1}^{\text{min}(Q,Q')-1}\left(\frac{u+\frac{2i}{h}(|Q'-Q|+2j)}{u-\frac{2i}{h}(|Q'-Q|+2j)}\right)^{2}~,~~~~~ \\
S^0_{(2Q|\a),(2Q'|\a)}(u) &=&\left(\frac{u+\frac{2i}{h}(Q'+Q)}{u-\frac{2i}{h}(Q'+Q)}\right) \left(\frac{u+\frac{2i}{h}|Q'-Q|}{u-\frac{2i}{h}|Q'-Q|} \right) \nn \\
&\times&
\prod_{j=1}^{\text{min}(Q,Q')-1}\left(\frac{u+\frac{2i}{h}(|Q'-Q|+2j)}{u-\frac{2i}{h}(|Q'-Q|+2j)}\right)^{2}~, \\
S^0_{(2Q-1|\a),(2Q'-1|\beta)}(u) &=& \left(\frac{u+\frac{2i}{h}(Q'+Q-1)}{u-\frac{2i}{h}(Q'+Q-1)} \right) \nn \\
&\times& \prod_{j=1}^{\text{min}(Q,Q')-1}\left(\frac{u+\frac{2i}{h}(|Q'-Q|-1+2j)}{u-\frac{2i}{h}(|Q'-Q|-1+2j)}\right)^{2}~,
\eea
\bea
S^0_{(2Q|\a),(2Q'|\beta)}(u)&=&\prod_{j=1}^{\text{min}(Q,Q')}
\left(\frac{u+\frac{2i}{h}(|Q'-Q|-1+2j)}{u-\frac{2i}{h}(|Q'-Q|-1+2j)}\right)^{2}~,
\\
S^0_{(2Q-1|\a),(2Q'|\a)}(u)&=&\left(S^0_{(2Q-1|\a'),(2Q'|\beta')}(-u)\right)^{-1}=\left(\frac{u+\frac{2i}{h}(Q+Q'-\frac{1}{2})}{u-\frac{2i}{h}(Q'-Q+\frac{1}{2})} \right)\nn\\
&\times&\prod_{j=1}^{Q-1}\left(\frac{u+\frac{2i}{h}(Q'-Q-\frac{1}{2}+2j)}{u-\frac{2i}{h}(Q'-Q+\frac{1}{2}+2j)}\right)^{2}~,\\
S^0_{(2Q|\a),(2Q'-1|\a)}(u)&=&\left(S^0_{(2Q|\a'),(2Q'-1|\beta')}(-u)\right)^{-1}=
\left(S^0_{(2Q'-1|\a''),(2Q|\a'')}(-u)\right)^{-1}\nn\\
&=&S^0_{(2Q'-1|\a'''),(2Q|\beta''')}(u)~.
\eea

Here we report the identities for the kernels  useful for the derivation of the Y-system.

\bea
\phi_{(\CQ'|\a),(\CQ|\a)}\left(z,u+\frac{i}{h}\right)+\phi_{(\CQ'|\a),(\CQ|\beta)}\left(z,u-\frac{i}{h}\right)&=&
\left(\phi_{(\CQ'|\a),(\CQ+1|\a)}+\phi_{(\CQ'|\a),(\CQ-1|\beta)}\right)(z,u)\nn\\
&-&\delta(z-u)\delta_{\CQ',\CQ+1}~,\\
\phi_{(\CQ'|\a),(\CQ|\a)}\left(z,u-\frac{i}{h}\right)+\phi_{(\CQ'|\a),(\CQ|\beta)}\left(z,u+\frac{i}{h}\right)&=&
\left(\phi_{(\CQ'|\a),(\CQ-1|\a)}+\phi_{(\CQ'|\a),(\CQ+1|\beta)}\right)(z,u)\nn\\
&-&\delta(z-u)\delta_{\CQ',\CQ-1}~,\\
\phi_{(y|-),\CQ}\left(z,u+\frac{i}{h}\right)+\phi_{(y|-),\CQ}\left(z,u-\frac{i}{h}\right)&=&
\sum_{\CQ'=1}^{\infty}I_{\CQ\CQ'}\phi_{(y|-),\CQ'}(z,u)  \nn \\
&+&\delta(z-u)\delta_{\CQ,1}~,\\
\phi_{(y|+),\CQ}\left(z,u+\frac{i}{h}\right)+\phi_{(y|+),\CQ}\left(z,u-\frac{i}{h}\right)&=&
\sum_{\CQ'=1}^{\infty}I_{\CQ\CQ'}\phi_{(y|+),\CQ'}(z,u)~,\\
\phi_{(v|M),\CQ}\left(z,u+\frac{i}{h}\right)+\phi_{(v|M),\CQ}\left(z,u-\frac{i}{h}\right)&=&
\sum_{\CQ'=1}^{\infty}I_{\CQ\CQ'}\phi_{(v|M),\CQ'}(z,u) \nn \\
&-&\delta(z-u)\delta_{\CQ-1,M}~,\\
\phi_{\CQ,(y|-)}\left(z,u+\frac{i}{h}\right)+\phi_{\CQ,(y|-)}\left(z,u-\frac{i}{h}\right)&=&
\phi_{\CQ,(v|1)}(z,u)+\delta(z-u)\delta_{\CQ,1}~,\\
\phi_{\CQ,(v|M)}\left(z,u+\frac{i}{h}\right)+\phi_{\CQ,(v|M)}\left(z,u-\frac{i}{h}\right)&=&
\sum_{M'=1}^{\infty}I_{MM'}\phi_{\CQ,(v,M')}(z,u) \nn \\
&+&\delta(z-u)\delta_{\CQ-1,M}~, \\
\phi_{KM} \left(u+\frac{i}{h} \right)+\phi_{KM}\left(u-\frac{i}{h} \right) &=&
\sum_{K'=1}^{\infty}  I_{K K'} \phi_{K'M}(u)+ I_{KM}\delta(u)~,\\
\phi_{M}\left(u+\frac{i}{h} \right)+\phi_{M}\left(u-\frac{i}{h}\right) &=&
 \phi_{M+1}(u)+\phi_{M-1}(u)+ \delta_{M,1}\delta(u) \nn \\
&=&  \phi_{M,1}(u)+ \delta_{M,1}\delta(u)~,
\eea

where $I_{KK'}=\delta_{K+1,K'}+\delta_{K-1,K'}$.
%
%
%


\end{document}